\documentclass{PoS}

\title{Precision measurements of the top quark couplings at the FCC}

\ShortTitle{Top-quark couplings at the FCC}

\author{\speaker{Patrick Janot}\thanks{On behalf of the FCC Design
    Study Group.}\\
        CERN, Geneva, Switzerland\\
        E-mail: \email{patrick.janot@cern.ch}}

\abstract{The design study of the Future Circular Colliders (FCC) in a
  100-km ring in the Geneva area has started at CERN at the beginning
  of 2014, as an option for post-LHC particle accelerators. The study
  has an emphasis on proton-proton and electron-positron high-energy
  frontier machines. In the current plans, the first step of the FCC
  physics programme would exploit a high-luminosity ${\rm e^+ e^-}$
  collider  called FCC-ee, with centre-of-mass energies ranging
  from below the Z pole to the ${\rm t\bar t}$ threshold and
  beyond, followed by 100\,TeV proton-proton collisions as ultimate
  goal. When combined, these two steps offer a large palette of complementary
  measurements and sensitivity for new physics. In particular, the
  association of the FCC-ee and the FCC-hh allows measurements of the
  top-quark electroweak and Yukawa couplings to be performed with
  unrivaled precision.}

\FullConference{The European Physical Society Conference on High Energy Physics\\
		22--29 July 2015\\
		Vienna, Austria}

\begin{document}

\parskip 0em
\section{The physics programme of the FCC-ee}
The core physics programme~\cite{FirstLook, AlainBlondel} of the
future high-luminosity ${\rm  e^+ e^-}$ circular collider (FCC-ee)
includes precision measurements at centre-of-mass energies from 90 to 
350\,GeV: \begin{itemize}
\setlength\itemsep{-0.3em}
\item[{\it (i)}] a scan of the Z pole~\cite{MogensDam}, enabling the
  measurements of, {\it e.g.},  $m_{\rm Z}$ and $\Gamma_{\rm Z}$ to
  better than 100\,keV,   $\sin^2\theta_W$ to $5\times 10^{-6}$,
  $\alpha_S(m_{\rm Z})$ to $2\times 10^{-4}$, $\alpha_{\rm
    QED}(m_{\rm   Z})$  to $2\times 10^{-5}$... , as well as a unique
  programme of rare
  process searches and flavour physics with up to $10^{13}$ Z
  decays~\cite{StephaneMonteil};  
\item[{\it (ii)}] a scan of the WW threshold~\cite{MogensDam} towards the
measurement of the W mass to 300\,keV;  
\item[{\it (iii)}] a Higgs factory~\cite{MarkusKlute}, for a tenfold
  improvement of the HL-LHC precision on the Higgs couplings, a
  measurement of the total width to 1\%, and a constraint on the invisible
  branching ratio at the per-mil level; 
\item[{\it (iv)}] a scan of the top threshold, for a measurement of the
top-quark mass~\cite{MogensDam} with a statistical precision of the
order of 10\,MeV. 
\end{itemize}

If these target precisions are reached and no
deviations with respect to the standard model predictions are
observed, a global fit to all these masurements~\cite{EllisYou} will set
constraints on weakly-coupled new physics up to a scale of $\sim
100$\,TeV, and on news physics coupled  to the Higgs sector up to
$\sim 10$\,TeV. This sensitivity to new physics matches  and even
extends  the direct new physics discovery potential of the FCC-hh.  

This fascinating perspective is made possible by the unique levels of
luminosity promised by circular colliders~\cite{MikeKoratzinos}
collected by up to four experiments.  The luminosities and the numbers of
events of a given type expected at each centre-of-mass energies for a
FCC-ee year of running with four interaction points are summarized in
Table~\ref{tab:lum}.  In this configuration, the core physics targets
of the FCC-ee programme would be met in 8 to 10 years at full
luminosity, after the commissioning period. 
\begin{table}[htbp]
\begin{center}
\begin{tabular}{|l|c|c|c|c|c|}
\hline $\sqrt{s}$ (GeV) & 90 & 160 & 240 & 350 & 350+ \\ 
\hline ${\cal L} {\rm (ab}^{-1}$/year) & 86.0 & 15.2 & 3.5 & 1.0 & 1.0 \\
\hline Events/year & $3.6\times 10^{12}$ & $6.1\times 10^{7}$ &
$7.0\times 10^{5}$ & $4.2\times 10^{5}$ & $2.5\times 10^{4}$ \\
\hline Event type & Z & WW & HZ & ${\rm t \bar t}$ & ${\rm WW \to H}$ \\
\hline\hline Years & 0.3 (2.5) & 1 & 3 & 0.5 & 3 \\ \hline 
\end{tabular}
\caption{Integrated luminosities in ${\rm ab}^{-1}$ and numbers of
  events expected for each year of running at the FCC-ee. The last row
  indicates the time needed to complete the core FCC-ee physics
  programme with $10^{12}$ ($10^{13}$) events at the Z pole. In this
  table, each year amounts to $10^7$ seconds of data taking.}
\label{tab:lum}
\end{center}
\end{table}
\vskip -1em
In this report, the FCC capability to measure the top-quark couplings is
investigated. In particular, the sensitivity of the FCC-ee to the top-quark
electroweak couplings is shown to be excellent for a centre-of-mass energy
just above the ${\rm t\bar t}$ threshold and without the benefit of
longitudinally polarized beams. The expected precision on these
couplings can be then used to extract the ttH coupling at the FCC-hh
with a per-cent-level precision. The FCC-ee targets are compared with
projections from the ILC in the ``H20'' design running
scenario~\cite{ILCH20}, as is done in Ref. ~\cite{ILCTOP}. This
  running scenario invokes a similar running time (about 8 to 10
  years),  albeit with a slightly more optimistic definition of a
``year'' ($1.6\times 10^7$ seconds of data taking) and a preliminary
modelling of the commissioning time.  

\section{The top-quark electroweak couplings at the FCC-ee}

The measurement of the top-quark electroweak couplings was, originally,
not part of the FCC-ee core physics programme. Indeed, this
measurement was claimed~\cite{ILCTDR} to require both a centre-of-mass
energy well beyond the top-pair production threshold and a large
longitudinal polarization of the incoming ${\rm e^\pm}$
beams. These claims were recently revisited~\cite{PatrickJanot} in the
context of the FCC-ee. 

In ${\rm e^+e^-}$ collisions, the availability of initial state
longitudinal polarization provides enhanced sensitivity to the
initial-state and, sometimes, final-state couplings to the photon and
the Z boson. Similar information can often be obtained, however, from
the final-state polarization, as is the case in the ${\rm e^+ e^-} \to
\gamma^\ast / {\rm Z} \to {\rm t \bar t}$ production at the FCC-ee:
in this process, anomalous couplings of the top quark to the Z and to
the photon would  alter the top-quark polarization.  This anomalous
polarization is  maximally transferred to the top-quark decay products
via the weak decay ${\rm t \to Wb}$, leading to a modification of the
final kinematics, and in particular of the angular and energy distributions
of the leptons from the W decays.  The large luminosity of the FCC-ee
allows in turn to extract precise measurements of the anomalous
couplings from these distributions. A similar situation was
encountered at LEP, where the measurement of the ${\rm 
  Z} \to \tau^+\tau^-$  event rate and of the $\tau$ polarization
sufficed to determine the couplings of the $\tau$ to the Z,
regardless of the initial state polarization. 

\begin{figure}
\begin{center}
\includegraphics[width=0.73\textwidth]{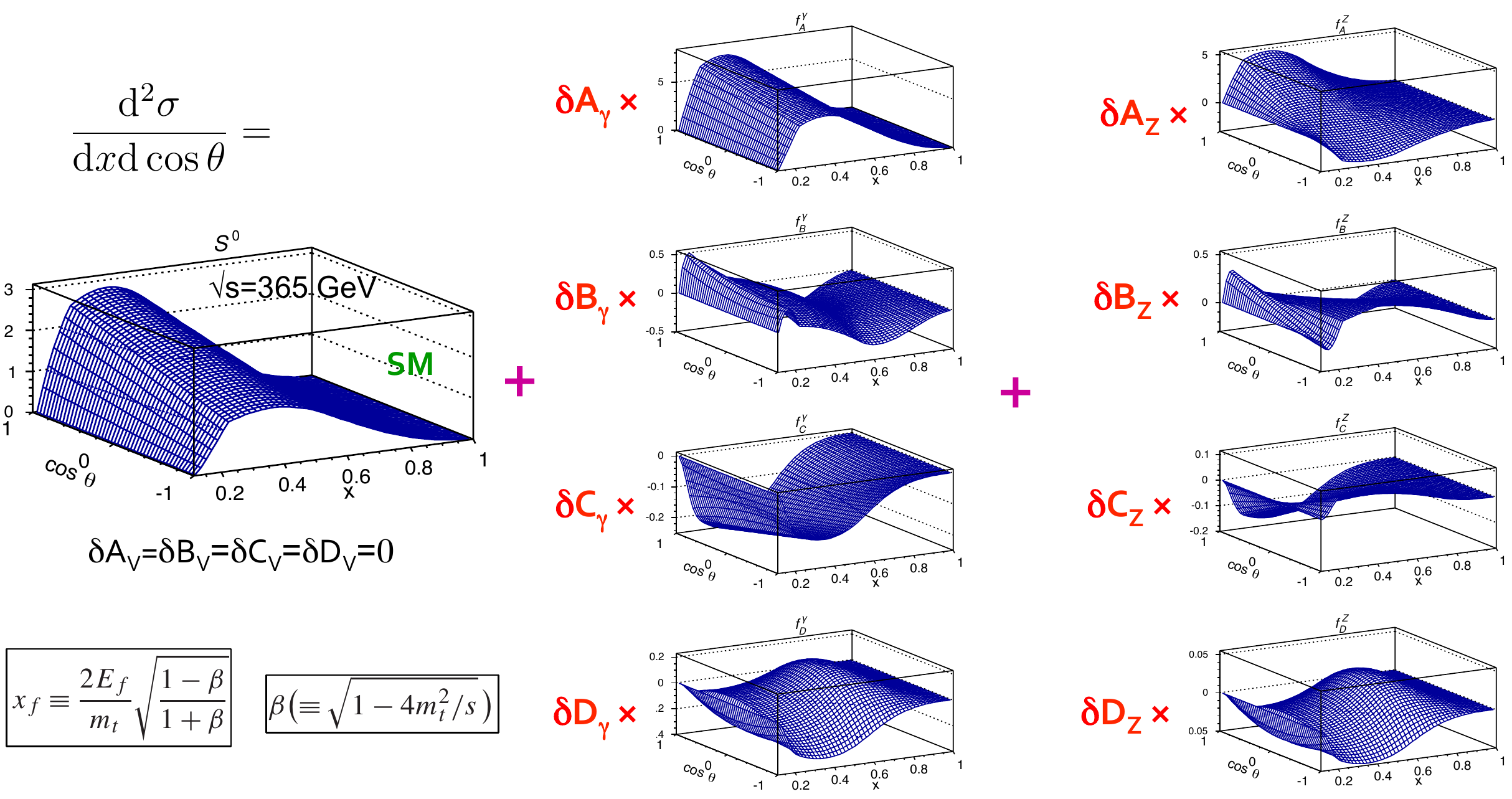}
\end{center}
\caption{The differential production cross section ${\rm d}^2\sigma /
  {\rm d}\cos\theta{\rm d}x_f$ for ${\rm e^+ e^-}
\to {\rm t \bar t} \to {\rm b \bar b q \bar q^\prime} \ell \nu_\ell$ 
($\ell = {\rm e}, \mu$) at $\sqrt{s} = 365$\,GeV, as a function of the
cosine of the lepton polar angle $\cos\theta$  and the reduced lepton
energy $x_f$ in the standard model (left). The additive modifications
arising from anomalous tt$\gamma$ and ttZ couplings are displayed on
the right.} 
\label{fig:lepton}
\end{figure}

The angular and energy distributions of the leptons from ${\rm e^+ e^-}
\to {\rm t \bar t} \to {\rm b \bar b q \bar q^\prime} \ell \nu_\ell$
($\ell = {\rm e}, \mu$) have been determined analytically in
Ref.~\cite{GrzadkowskiHioki} as a function of the centre-of-mass
energy and of the incoming beam polarization, in presence of anomalous
tt$\gamma$ and ttZ couplings, denoted $\delta A_{\gamma,{\rm Z}}$,
$\delta B_{\gamma,{\rm Z}}$,  $\delta C_{\gamma,{\rm Z}}$, and
$\delta D_{\gamma,{\rm Z}}$. These anomalous couplings modify the ttV
vertex ($V =\gamma$, Z) as follows:
\begin{equation}
\Gamma^\mu_{ttV} = {g\over 2} \left[ \gamma^\mu \left\{ (A_V+\delta A_V) - 
\gamma_5 (B_V+\delta B_V) \right\} + 
{(p_t -p_{\bar t})^\mu\over 2 m_{\rm t}}  
\left( \delta C_V - \delta D_V \gamma_5 \right) \right],
\end{equation}
where $A_V$ and $B_V$ are the vector and axial standard model
top couplings, and modify the lepton angular and energy
distribtution as sketched in Fig.~\ref{fig:lepton}. 

It is shown in Ref.~\cite{PatrickJanot} that a likelihood fit of the
anomalous couplings to the double differential cross section is
statistically optimal at the FCC-ee for $\sqrt{s} = 365$\,GeV, without
beam polarization. With conservative assumptions on
lepton identification, b-tagging efficiencies, and lepton angular
and momentum resolutions, it is estimated therein that absolute
precisions of the order of  $1 [3] \times 10^{-3}$ ($1 [2] \times
10^{-2}$) can be expected after three FCC-ee years at $\sqrt{s} =
365$\,GeV on $\delta A_{\gamma  [\rm Z]}$ ($\delta B_{\gamma [\rm Z]}$), if
$\delta C_{\gamma  [\rm Z]} = 0$. A full  simulation of the CLIC/ILD
detector operated in the FCC-ee environment~\cite{NicoloTommaso}  has
recently confirmed these analytical preliminary estimates.  

These target precisions will be met only if the theoretical prediction
for the top-pair cross section can be kept under control to better
than $\pm 2\%$, perhaps a serious challenge only 20\,GeV above the
production threshold. It is inferred in Ref.~\cite{JRReuters}, however,
that the total theoretical uncertainty is, already today, at the level
of $\pm 4\%$ at $\sqrt{s} = 365$\,GeV. A factor 2 improvement might be
beyond the current state of the art, but is probably within reach on
the time scale required by the FCC-ee.  

\begin{figure}[htbp]
\begin{center}
\includegraphics[width=0.48\textwidth]{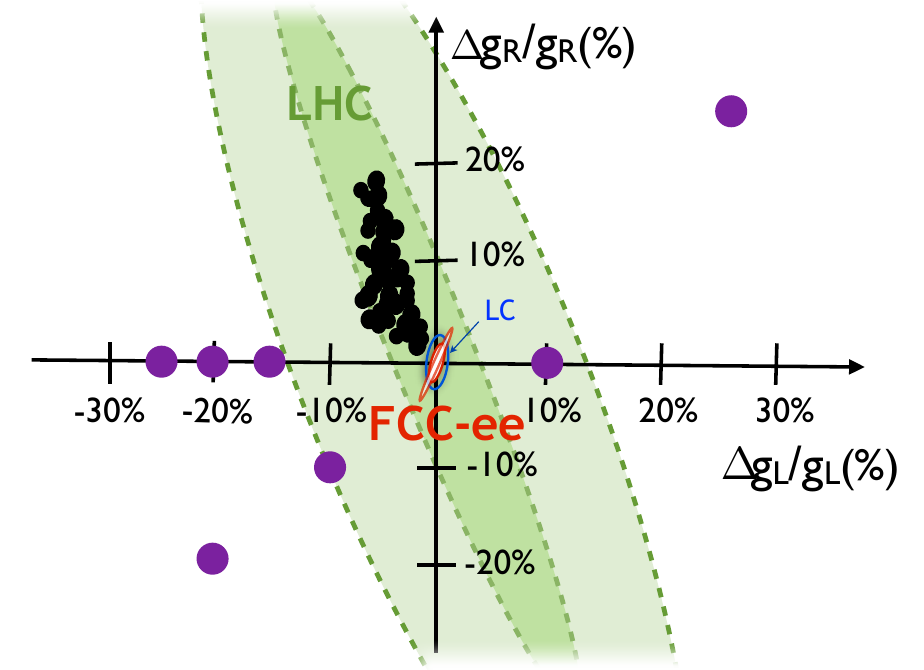}
\includegraphics[width=0.42\textwidth]{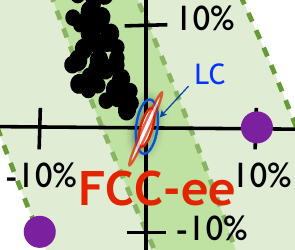}
\end{center}
\caption{(Adapted from Ref.~\cite{StefaniaDeCurtis}) Expected relative
  precision on the ${\rm Zt_Lt_L}$ and ${\rm Zt_Rt_R}$ couplings at
  the LHC (lighter green), the HL-LHC (darker  green), the ILC (blue)
  and the FCC-ee (orange, red). The left plot is a zoom of the right
  plot in the $\pm 10\%$ window. Typical deviations from the standard
  model in various new physics models are represented by the purple
  dots. The black dots indicate the deviations expected for different
  parameter choices of 4D  Composite Higgs Models, with $f<2$\,TeV. 
  For the FCC-ee, the
  orange ellipse is  obtained from an analysis of the lepton angular and energy
  distributions, while the smaller red ellipse is obtained if the angular and
  energy distributions of the b jets can be  exploited.}
\label{fig:grgl}
\end{figure}

The sensitivity of these projections to new physics is evaluated
under the mild additional assumption that the electric charge of the top
quark is $+2/3$. New physics information is then entirely contained in
the ttZ axial and vector couplings $A_{\rm Z}$ and $B_{\rm Z}$ or,
equivalently ( as is done for example in Ref.~\cite{StefaniaDeCurtis}),
in the couplings to the left-handed and right-handed top quark,  $g_L$
and $g_R$, trivially related to $A_{\rm Z}$ and $B_{\rm Z}$:  
\begin{equation}
g_R+g_L = gA_{\rm Z} {\rm \ \ \ and \ \ \ } g_R-g_L = gB_{\rm Z} ,
\end{equation} 
where $g$ is the weak coupling constant. The relative precision expected at
FCC-ee for $g_L$ and $g_R$ is displayed in Fig.~\ref{fig:grgl}, and is
compared to the projections made for the LHC, the
HL-LHC~\cite{RontschSchulze} and the ILC~\cite{ILCTOP, Poeschl}. 
The most exotic new physics models -- represented by purple dots in
Fig.~\ref{fig:grgl} -- will already be explored by the modest precision
expected at the HL-LHC, but an ${\rm e^+ e^-}$ collider will be needed
to disentangle the standard model from, {\it e.g.}, 4D composite Higgs models
(4DCHM, black dots). If the characteristic scale of these models, $f$, is
constrained to be smaller than 2\,TeV,  as is the case in
Fig.~\ref{fig:grgl}, the FCC-ee will be able to discover them from the
sole analysis of the lepton angular and energy distributions in
top-quark decays with a $5\sigma$ significance. The significance
expected at the ILC, from the analysis of the top forward-backward
asymmetry and the total rate, with two different beam polarization
configurations (${\cal P}_- = \pm 0.8$ and ${\cal P}_+ = \mp 0.3$),
would be limited to $1.5\sigma$.  

\begin{figure}[htbp]
\begin{center}
\includegraphics[width=0.60\textwidth]{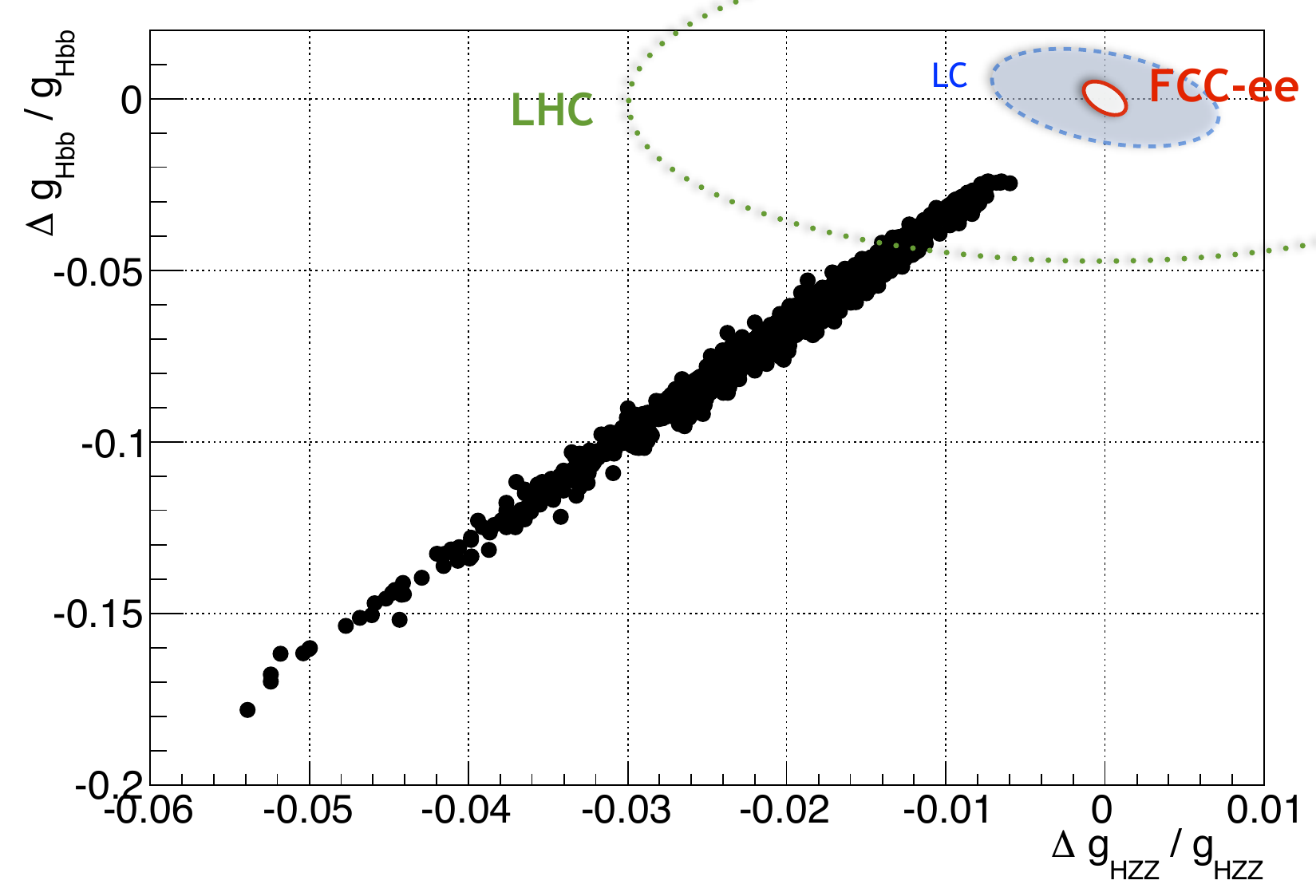}
\end{center}
\caption{Expected relative precision on the HZZ and Hbb couplings at 
  the HL-LHC (dotted green), the ILC (dashed blue) and the FCC-ee
  (full red). The black dots indicate the deviations expected for the
  same parameter choices of 4D  Composite Higgs Models ($f<2$\,TeV) as in
  Fig.~2.} 
\label{fig:higgs}
\end{figure}

It is important to realize, however, that top-quark couplings might
not be the only quantities affected by new physics. Composite Higgs
models, in particular, affect Higgs couplings as well. The deviations
expected on the Higgs couplings to the Z boson and to the b quark for
the same set of 4DHCM as in Fig.~\ref{fig:grgl} are displayed in
Fig.~\ref{fig:higgs}, and compared to the corresponding relative
precisions expected at the HL-LHC~\cite{MarkusKlute}, the
ILC~\cite{ILCPhysics} and the FCC-ee~\cite{FirstLook}.  From these
Higgs coupling measurements, the FCC-ee will be able to discover these
models with a 10$\sigma$  significance (to be compared to about
2$\sigma$ for the ILC), complementary to, and possibly earlier than, the
top electroweak coupling measurements. 

\section{The top-quark Yukawa couping at the FCC-hh}
New physics is also expected to show up in top-quark Yukawa coupling
to the Higgs boson. For the previous set of composite Higgs models,
deviations similar to or slightly smaller than those of the b-quark
Yukawa couplings are expected, typically of the order of 1 to 2\% for
$f = 2$\,TeV. It has been shown~\cite{Snowmass} that ${\rm e^+ e^-}$
colliders with centre-of-mass energy significantly larger than the
FCC-ee or the ILC are needed to measure the top-quark Yukawa couplings
with a precision significantly smaller than 10\%: for $\sqrt{s} = 1$ to
$3$\,TeV, a precision of 2 to 4\% can be contemplated, which still
does not suffice to challenge these models. As shown in
Table~\ref{tab:ttH}, a similar performance is expected from the
HL-LHC, with $3\,{\rm ab}^{-1}$ at $\sqrt{s} = 13$\,TeV.  

The FCC opens a unique window for this measurement, with a combination
of the FCC-ee and the FCC-hh data~\cite{MLM}. In pp collisions, the
${\rm pp \to  t\bar t H}$ cross section increases by a factor 70 --
from 0.47 to 33.2\,pb -- when the centre-of-mass energy goes from 13
to 100\,TeV. With a target integrated luminosity of $20\,{\rm
  ab}^{-1}$, almost a billion ${\rm t\bar t H}$ events are expected to
be produced at the FCC-hh, leading to a negligible statistical
uncertainty on the top Yukawa coupling. The measurement of the ${\rm
  t\bar t H}$ cross section would therefore be only limited by
systematic uncertainties, dominated by the
renormalization/factorization scale and PDF uncertainties at the level
of $\pm 10\%$. These uncertainties almost exactly cancel in the ratio
$\sigma({\rm t\bar t  H})/\sigma({\rm t\bar t Z}) \simeq 0.6$ to
better than $\pm 2\%$ at $\sqrt{s} = 100$\,TeV. A measurement of this
ratio amounts to measuring the quantity $R$ defined by 
\begin{equation}
\label{eq:R}
R = \frac{\lambda_{\rm t}^2}{g_R^2 + g_L^2} \times \frac{{\rm BR(H \to
  b\bar b, ZZ,}, \tau^+\tau^-, \dots{\rm )}}{{\rm BR(Z} \to
\ell^+\ell^-{\rm )}},
\end{equation}
where $\lambda_{\rm t}$ is the top-quark Yukawa coupling and $g_{L,R}$
are the top-quark weak couplings to the Z boson.  The first ratio of
Eq.~\ref{eq:R} can be predicted, already today, with an accuracy
better than 2\%, and the FCC-ee can deliver a measurement of $g_L^2 +
g_R^2$ with a precision of $\sim 2\%$, as can be deduced from
Fig.~\ref{fig:grgl}. For all practical purposes, the second ratio is
perfectly known, as the Higgs branching fractions can be measured in a
model-independent way at the FCC-ee with a sub-per-cent precision, and 
the Z leptonic branching ratios will be known with a precision better
than $10^{-5}$.  A combination of all these measurements therefore
allows the value of top-quark Yukawa coupling to be inferred with 
an accuracy of $\sim 1.5\%$ at the FCC. 

These projections are summarized in Table~\ref{tab:ttH}, both for the
top-quark Yukawa and for the Higgs trilinear
self-coupling~\cite{MarkusKlute}, another important parameter to be
measured at future colliders.
\begin{table}[htbp]
\begin{center}
\begin{tabular}{|c|c|c|c|c|}
\hline Collider & HL-LHC & ILC & LC 1-3\,TeV & FCC-ee+hh\\ 
\hline $\lambda_{\rm t}$ & $4\%$ & $14\%$ & $2-4\%$ & $1-2\%$  \\
\hline $\lambda_{\rm H}$ & $50\%$ & $83\%$ & $10-15\%$ & $5-10\%$  \\ \hline 
\end{tabular}
\caption{Expected precision on the top-quark Yukawa coupling (first
  row) and, for completeness, on the Higgs trilinear self-coupling
  (second row) at the HL-LHC, the ILC, a linear ${\rm e^+ e^-}$
  collider with $\sqrt{s} = 1$ to $3$\,TeV, and the FCC -- with a
  combination of the FCC-ee and the FCC-hh data. }
\label{tab:ttH}
\end{center}
\end{table}

\section{Conclusion}

The measurement of the angular and energy distributions in
semi-leptonic ${\rm t\bar t}$ events at the FCC-ee has a strong
potential for a precise determination of the top-quark  electroweak
couplings. The optimal centre-of-mass energy for this measurement is
just above the ${\rm t\bar t}$ threshold, typically $\sqrt{s} =
365$\,GeV.  The lack of beam polarization is compensated
by the polarization of the top quarks, and by a large integrated
luminosity. In combination with this measurement and those of the
Higgs boson branching fractions at the FCC-ee, the large ${\rm t\bar tH}$ and ${\rm
  t\bar tZ}$ production cross sections  at the FCC-hh, as well as the
cancellation of the dominant theory uncertainties in their ratio,
allow the top-quark Yukawa coupling to be measured with a per-cent
level precision. 

Taken in isolation, these measurements at the FCC are already more
sensitive to the presence of new physics than those at other collider
options considered so far. New physics responsible for anomalous top
couplings is also likely to affect the properties of the Z, W, and
Higgs bosons. The FCC will therefore also have many opportunities for
new physics discoveries, as well as a palette of precision measurements
to help identify the underlying theory, through a global fit to these
properties and their correlations.  

The combination of the FCC-ee and the FCC-hh offers, for a great cost
effectiveness, the best precision and the best search reach of all
options presently on the market.

\end{document}